\documentclass[aps,prb,twocolumn,floatfix,floats,showpacs,superscriptaddress]{revtex4}
\usepackage{graphicx,latexsym}
\usepackage{dcolumn}
\usepackage{amsmath}

\def\fig#1{Fig.~\ref{#1}}

\def\be{\begin{equation}}
\def\ee{\end{equation}}
\def\bea{\begin{eqnarray}}
\def\eea{\end{eqnarray}}

\begin{document}

\title{Tuning of coupling modes in laterally parallel double open quantum dots}

\author{Chi-Shung Tang}
\affiliation{Physics Division, National Center for Theoretical
        Sciences, P.O.\ Box 2-131, Hsinchu 30013, Taiwan}
\author{Wing Wa Yu}
\affiliation{Science Institute, University of Iceland,
        Dunhaga 3, IS-107 Reykjavik, Iceland}
\author{Vidar Gudmundsson}
\affiliation{Science Institute, University of Iceland,
        Dunhaga 3, IS-107 Reykjavik, Iceland}

%

\begin{abstract}
We consider electronic transport through laterally parallel double
open quantum dots embedded in a quantum wire in a perpendicular
magnetic field. The coupling modes of the dots are tunable by
adjusting the strength of a central barrier and the applied magnetic
field. Probability density and electron current density are
calculated to demonstrate transport effects including magnetic
blocking, magnetic turbulence, and a hole-like quasibound state
feature. Fano to dip line-shape crossover in the conductance is
found by varying the magnetic field.
\end{abstract}

\pacs{73.23.-b, 73.21.La, 73.21.Hb, 85.35.Ds}


\maketitle

%
%

\section{introduction}

Electronic transport through an open quantum dot has attracted broad
attention~\cite{Chang94,Chan95,Persson95,Keller96,Wang96,Bird,Zoz98,Tang03,
Clerk01,Vavilov05,Kim03,Mendoza05} due to its potential in the
investigation of various bound-state features,~\cite{Tang03} phase
coherence,~\cite{Clerk01,Vavilov05} and wave function
imaging.~\cite{Kim03,Mendoza05} In high electron mobility samples at
low temperatures, the electron phase coherent length may be longer
than the dimension of the open dot system, allowing electrons to
remain coherent while traversing the system with negligible impurity
effects. Moreover, since the outgoing electrons from the open dot
are strongly coupled to reservoirs without tunneling, the Coulomb
effects are negligible.

By coupling two quantum dots in series or in parallel, a double
quantum dot is formed.~\cite{vdWiel03,Chen04,Lu05,Kie05} Quantum
transport through such a double dot system has attracted
considerable attention due to its versatility for various
applications.~\cite{Oosterkamp98,Jeong01,Kostyrko05,Zutic04,Requist05,
Johnson05,Jouravlev05,Kuzmenko02,Hartmann04,Ding05} The double dot
system provides possible new mechanism compared to a single quantum
dot as electrons could be coupled between the two dots, thus forming
an artificial quantum dot molecular
junction.~\cite{Oosterkamp98,Jeong01,Kostyrko05}  In addition, the
coupled dot system is likely to be important in quantum information
processing,~\cite{Zutic04,Requist05,Johnson05,Jouravlev05} where
external field manipulation and quantum coherence are both required.
Thus far these coupled dot systems are, however, assumed to be
isolated and can be described by an Anderson-type
model.~\cite{Kuzmenko02,Hartmann04,Ding05}

We would like to emphasize that the adiabaticity of the dot-lead
connection holds only for large quantum dots. As the dot size
shrinks and approaches the realm of the Fermi wave length, the
dot-lead connection no longer remains adiabatic. Experimental
findings in electronic transport through quantum dots, such as the
individual eigenstates of isolated dot~\cite{Zozou99} and the
recurrence of specific groups of wave function scars in the
dot,~\cite{Bird} indicate unequivocally the mode-mixed scattering at
the dot-lead connections.  It has been shown that the embedded
quantum structures can lead to a complicated mode
mixing.~\cite{Gudmundsson05:BT,Gudny05}

\begin{figure}[b]
      \includegraphics[width=0.45\textwidth,angle=0]{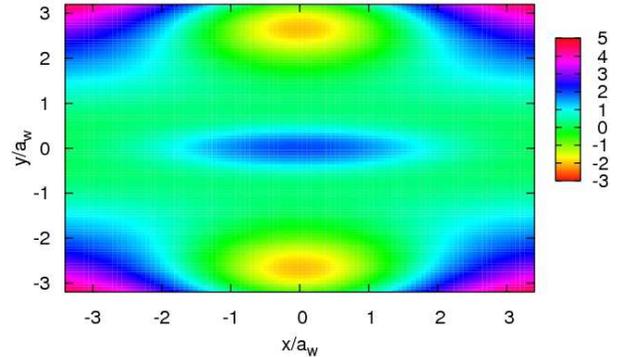}
      \caption{Schematic illustration of the laterally parallel DOQD system
      containing two open quantum dots (yellow) to the sides
      and a central barrier (blue).
      The color scale on the right shows the potential height in meV.
      The parameters are $a_w = 33.7$ nm, $\hbar\Omega_0 = 1.0$ meV,
      $V_1 = V_3 = -6.0$ meV, and $V_2 = 2.0$ meV.}
      \label{system}
\end{figure}

In this paper we study the tuning of coupling modes of parallel
double open quantum dots (DOQDs) embedded in a quantum wire. The
mode-coupling in this system is coherently adjusted by a central
elongated potential barrier separating the system into upper and
lower channels, as depicted in \fig{system}.  In addition, an
external perpendicular magnetic field is applied to manipulate the
electronic cyclotron motion and the coupling between the upper dot
(UD) and the lower dot (LD). It is important to note that since the
DOQD system is strongly coupled to the source and drain reservoirs,
the quantum interference effects are strong and cannot be treated
like an isolated dot using Anderson-type model or solving the rate
equation for the Fock-Darwin spectrum.~\cite{Rogge05} Here we employ
Lippmann-Schwinger approach~\cite{Gudmundsson05:BT,Gudny05} that
allows us to handle a wide range of smooth scatterers embedded in a
wire and access the electron probability distribution as well as the
electron current flow in the system.

One robust transport phenomenon in open quantum structures is the
quasibound-state feature with positive~\cite{Chu89,Faist90} or
negative~\cite{Laughlin83,Vidar05} binding energy. Indeed, the
transport properties of a wire with either static or time-dependent
scatters can exhibit significant quasibound-state
features.~\cite{Chu89,Faist90,Laughlin83,Vidar05,TangChu} In a
laterally parallel DOQD system, operating the central barrier
simultaneously adjusts the dot-dot and dot-lead coupling and tilts
the potential of the side dots to affect the alignment of the
electron energy with the discrete levels in the dots. The electronic
transport thus manifests different types of resonant features.

%
\section{model: a Lippmann-Schwinger approach}

The system under investigation is composed of a laterally parallel
DOQD embedded in a quantum wire with confining potential $V_{\rm
c}(y) = \frac{1}{2} m^*\Omega_0^2 y^2$, and hence the electrons are
transported through the wire with characteristic energy scale
$\hbar\Omega_0$ in the transverse direction. The parallel double
open quantum dots are separated by a central barrier that can be
used to tune the dot coupling and adjust the mode mixing between the
wire subbands and the dot levels. The electrons incident from the
left reservoir impinge on the parallel DOQD system modeled by the
Gaussian-type scattering potential
\begin{equation}
V_{\rm sc}(x,y) = \sum_{i=1}^3 V_{i} \exp { \left[ -\alpha x^2 -
\beta_{i} \left( y-y_{i} \right)^2 \right]}.
\end{equation}
Here $V_1$ and $V_3$ are negative indicating, respectively, the
depth of the UD and the LD, and $V_2$ is positive describing the
height of the central barrier. The three potentials have the same
length (same $\alpha$) while the central barrier is a little
narrower then the two dots ($\beta_1 = \beta_3 < \beta_2$) so that
the UD is allowed to couple with the LD. In a perpendicular magnetic
field ${\bf B} = B {\hat{\bf z}}$, the parabolic confinement defines
an effective magnetic length $a_w= (\hbar/m^*\Omega_w)^{1/2}$ in the
wire and the subband energy levels
\begin{equation}
E_n = \left(n + \frac{1}{2}\right)\hbar\Omega_w + \frac{\left( k a_w
\right)^2}{2} \frac{\left( \hbar\Omega_0 \right)^2}{\hbar\Omega_w}
\, ,
\end{equation}
where $k$ is the magnitude of the electron wave vector along the
wire and $n=0,1,\cdots$.  This defines the effective subband
separation $\hbar \Omega_w$, where $\Omega_w^2 = \omega_{\rm c}^2 +
\Omega_0^2$ is related to the cyclotron frequency $\omega_{\rm
c}=eB/(m^*c)$ and the characteristic frequency $\Omega_0$ for the
parabolic model.  We assume a value of $m^* = 0.067 m$ for the
effective mass of an electron in GaAs-based material.

In the following we employ a mixed momentum-coordinate
presentation~\cite{Gurvitz95}
\begin{equation}
\Psi_E(p,y)\equiv\int dx\: e^{-ipx} \psi_E(x,y)
 \end{equation}
and utilize a channel-mode expansion
$\Psi_E(p,y)=\sum_n\varphi_n(p)\phi_n(p,y)$
 to obtain a set of coupled Lippmann-Schwinger integral equations in
the momentum space.  As such, these equations can then be
transformed into integral equations for the $T$-matrix to facilitate
numerical calculation.~\cite{Gudmundsson05:BT}  According to the
Landauer-B{\"u}ttiker formalism the energy dependence of the
conductance can be calculated~\cite{Gudmundsson05:BT}
\begin{equation}
G(E) = G_0 {\rm Tr}\left[{\bf t}^{\dag}(E){\bf t}(E)\right]
\end{equation}
with the conductance quantum $G_0=2e^2/h$.  Furthermore, with the
scattering wave function obtained from the $T$-matrix we calculate
in configuration space the probability density $|\psi_E(x,y)|^2$ and
the electron current density
\begin{equation}
{\bf J} = \frac{e}{m} {\rm Re}\left[\psi_E^* \left({\bf p} +
\frac{e}{c}{\bf A}\right)\psi_E\right]
\end{equation}
to clearly demonstrate the transport mechanism and provide detailed
insight into the coupling nature of the parallel DOQD system.  Here
$-e$ is the charge of an electron.  For embedded wire systems it is
convenient to choose the vector potential ${\bf A} = (-By,0,0)$ in a
Landau gauge.

\section{results and discussion}

To study the transport behavior in a laterally parallel DOQD system,
we consider a broad parabolic wire with confinement energy
$\hbar\Omega_0 = 1.0$ meV.  This energy corresponds to $a_w = 33.7$
nm in zero magnetic field.  In the DOQD system, the central barrier
and the side dots have the same effective length $L \simeq 141$ nm
($\alpha = 2\times 10^{-4}$ nm$^{-2}$), while the width of the
central barrier ($\beta_2 = 4\times 10^{-3}$ nm$^{-2}$) is narrower
than the side dots ($\beta_1=\beta_3 = 0.7\times 10^{-3}$ nm$^{-2}$)
to facilitate the UD-LD coupling.  By choosing these parameters, the
effective width of the central barrier and the side dots are,
respectively, $W_{\rm CB}\simeq 32$ nm and $W_{\rm SD}\simeq 76$ nm.
The two open dots have the same potential depth $V_1 = V_3 = -6.0$
meV and are separated by 100 nm from the central barrier ($y_2 =
0$), namely, $y_1 = -y_3 = 100$ nm.

In performing the numerical calculation, a sufficient total number
of quantum channels including evanescent modes in momentum space is
needed to satisfy the conservation of current condition. Numerical
accuracy is assured by comparing the data obtained from a larger
basis set.  In order to investigate the characteristics of the
energy-dependent conductance $G(E)$ by tuning the strength of a
central barrier and the applied magnetic field, below we show the
conductance as a function of
\begin{equation}
X = \frac{E}{\hbar\Omega_w} + \frac{1}{2}\, .
\end{equation}
The integer part of the parameter $X$ counts how many propagating
channels in the wire are open for an incoming electron with energy
$E$.

\subsection{Tuning the central barrier}

\begin{figure}[t]
      \includegraphics[width=0.48\textwidth,angle=0]{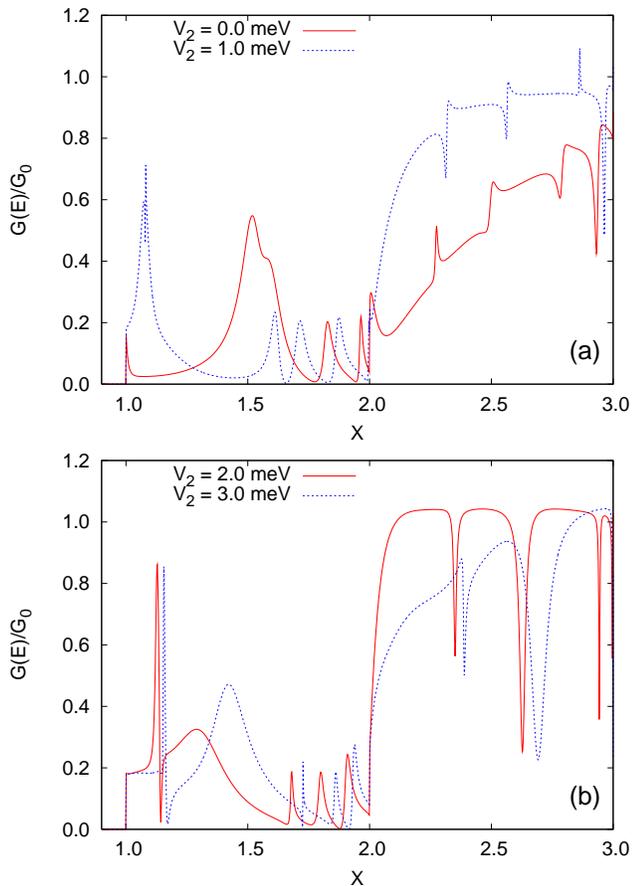}
      \caption{The conductance in units of $G_0=2e^2/h$
               as a function of $X$ in a magnetic field $B = 0.5$ T
               with different height of the central barrier $V_2$ = (a) 0.0
               (solid),
               and 1.0 (dashed) meV; (b) 2.0 (solid), and 3.0 (dashed) meV.
               Other parameters are
               $V_1=V_3=-6.0$ meV and $L=4.82 a_w$.
               } \label{GEV}
\end{figure}
In \fig{GEV}, we investigate the tuning effects of the central
barrier $V_2$ in a magnetic field $B = 0.5$ T. Correspondingly, the
effective magnetic length $a_w = 29.34$ nm and the effective subband
separation $\hbar\Omega_w = 1.32$ meV, and hence $a_w = a_{\rm
2D}\sqrt{\omega_{\rm c}/\Omega_w}\simeq 0.8 a_{\rm 2D}$ with $a_{\rm
2D}$ being the magnetic length in a flat two-dimensional system. The
center of the side dots are located at $(x_c,y_c) = (0, \pm 3.4
a_w)$.  Since the magnetic field at 0.5 Tesla provides an optimal
condition to facilitate the UD-LD coupling and enhance a cyclotron
backscattering, the general features in $G$ are strongly suppressed
due to a {\it magnetic blocking} effect. In the single-mode regime
($1<X<2$), this magnetic blocking effect becomes significant in the
high kinetic energy (KE) regime. In the double-mode regime
($2<X<3$), if the height of the central barrier is close to
$\hbar\Omega_w$ (such as $V_2 = 1.0$ and $2.0$ meV) [see \fig{GEV}],
the magnetic blocking effect would be suppressed and the conductance
envelop manifests a plateau-like structure at $G/G_0\approx 1.0$.
\begin{figure}[t]
      \includegraphics[width=0.45\textwidth,angle=0]{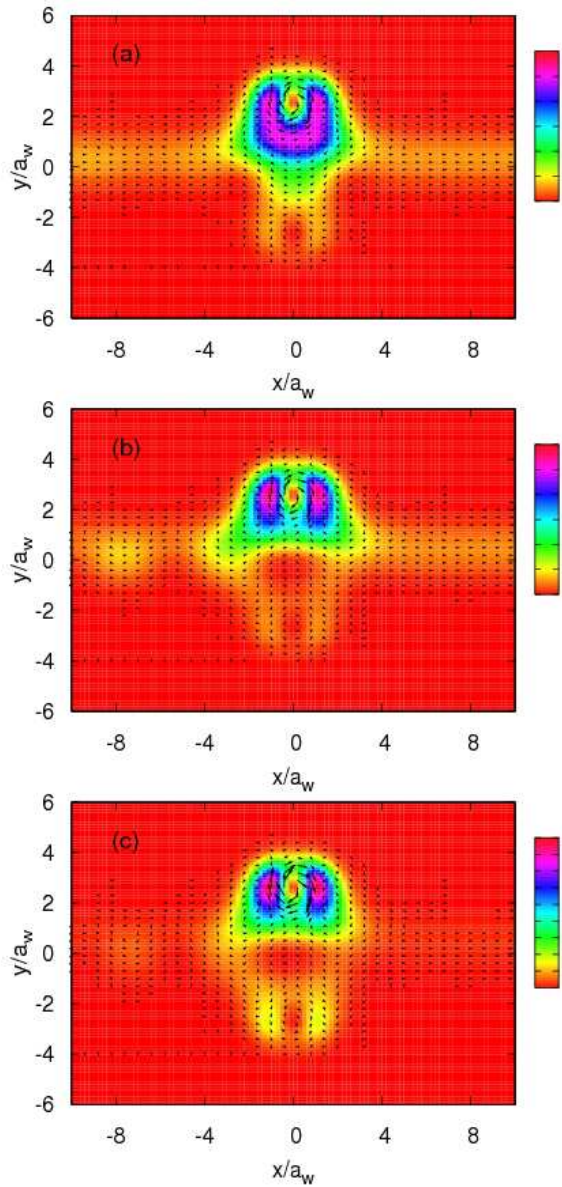}
      \caption{Probability density and electron current density (black arrows)
       are plotted as functions of $x$ and $y$  in a magnetic field $B = 0.5$ T
               with different strength of the central barrier in the low-KE single-mode
               regime: (a)$V_2 = 1.0$ meV at $X = 1.081$; (b) $V_2 = 2.0$ meV at $X = 1.143$;
               and (c) $V_2 = 3.0$ meV at $X = 1.171$. Other parameters are
               $V_1=V_3=-6.0$ meV and $L=4.82 a_w$.
               } \label{V123X1L}
\end{figure}

In the low-KE single-mode regime, we find sharp Fano-type
peaks~\cite{Fano61} in $G$: They are at $X=1.081$ for $V_2 =1.0$
meV, $X=1.129$ for $V_2=2.0$ meV, and $X=1.155$ for $V_2=3.0$ meV.
These structures represent resonant transmission of electrons with
continuous subband energy in the leads that strongly couple to the
discrete level in the UD forming a long-lived quasibound state with
quantum number $(n_x,n_y)=(1,2)$ in the UD.~\cite{Tang03}  It is
important to point out that the clockwise cyclotron motion in the UD
indicates a {\it magnetic hole-like state}. These states are shifted
by changing $V_2$ because of the tilting effect to the dot potential
by the central barrier. Moreover, the Fano peak with $G/G_0=0.71$ at
$X=1.081$ for the case of $V_2=1.0$ meV is actually a weak
dip-and-peak structure.  This weaker structure imposed on the peak
is due to the lower central barrier such that the electrons may have
stronger coupling to the LD.  When the central barrier is tuned to
be higher, e.g. $2.0$ or $3.0$ meV, a perfect (2,1)-like quasibound
state can be constructed in the UD, and the conductance manifests
strong peak-and-dip structures. On the other hand, there are Fano
dips at $X=1.079$ for $V_2=1.0$ meV, $X=1.143$ for $V_2=2.0$ meV,
and $X=1.171$ for $V_2=3.0$ meV. These structures correspond to
electrons coupled with higher probability to the LD forming a
hole-like state, and then being backscattered to the left lead
resulting in a conductance dip.
\begin{figure}[t]
      \includegraphics[width=0.45\textwidth,angle=0]{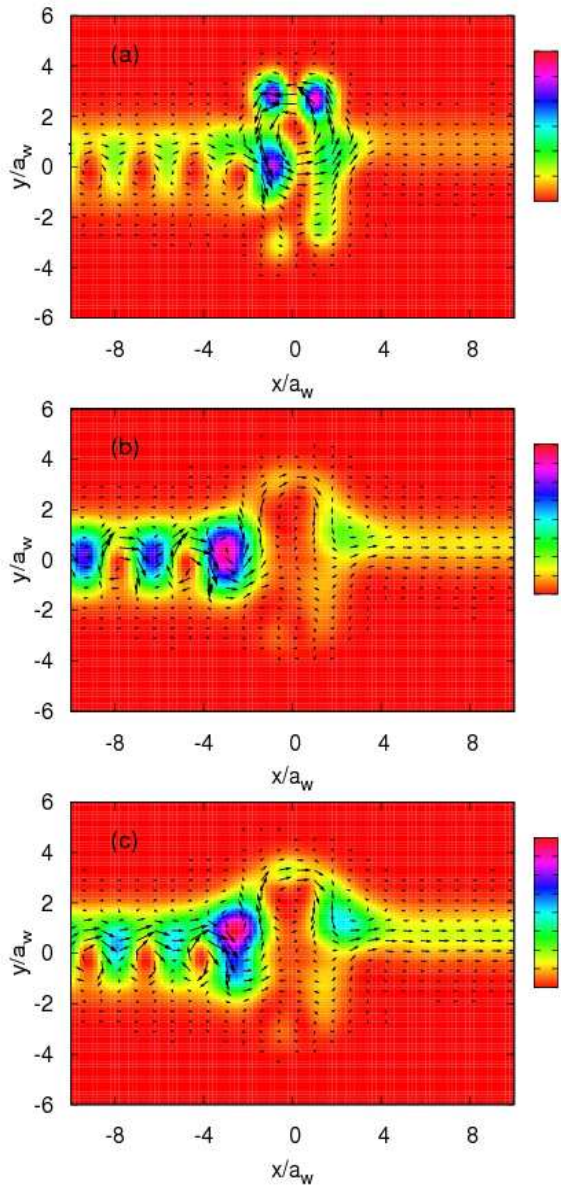}
      \caption{Probability density and electron current density (black arrows)
       are plotted as functions of $x$ and $y$  in a magnetic field $B = 0.5$ T
               with different strength of the central barrier in the mediate-KE
               single-mode regime: (a) $V_2 = 0.0$ meV at $X = 1.52$;
               (b) $V_2 = 2.0$ meV at $X = 1.29$;
               and (c) $V_2 = 3.0$ meV at $X = 1.42$. Other parameters are
               $V_1=V_3=-6.0$ meV and $L=4.82 a_w$.
               } \label{V023X1M}
\end{figure}

\begin{figure}[t]
      \includegraphics[width=0.45\textwidth,angle=0]{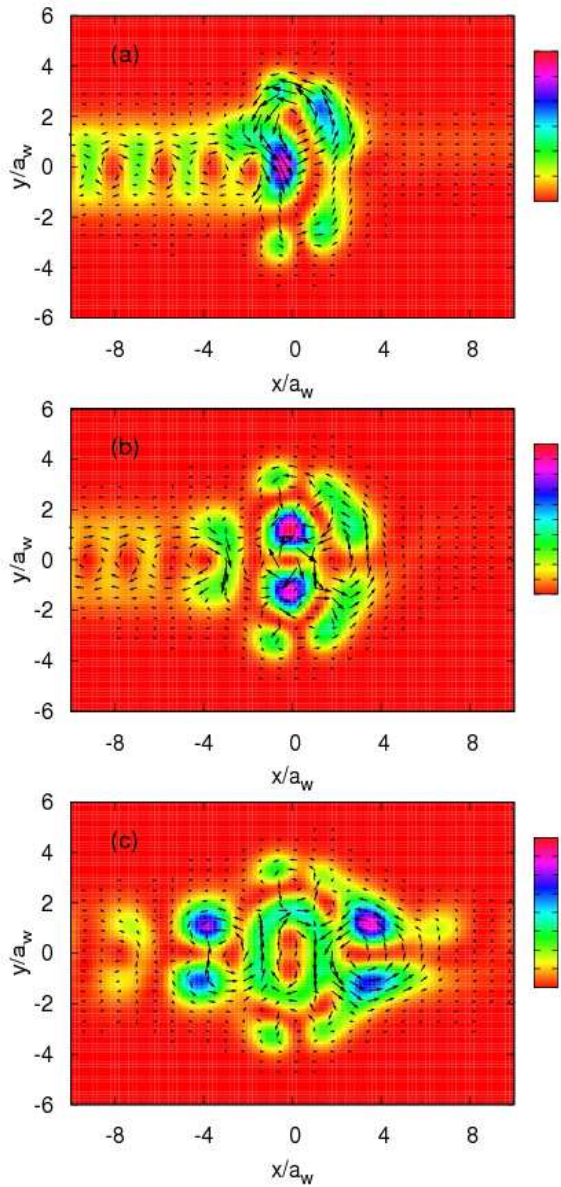}
      \caption{Probability density and electron current density (black arrows)
         for the case of $V_2 = 0$ meV and $B = 0.5$ T at $X =$ (a) 1.58; (b) 1.83;
         and (c) 1.97. Other parameters are $V_1=V_3=-6.0$ meV and $L=4.82 a_w$.
               } \label{V0X1B5}
\end{figure}

In the mediate-KE single-mode regime, there is a hump structure in
$G$ except for the case of $V_2 = 1.0$ meV. The broad nature of
these structures implies the short electron dwell time of the
quasibound states in the UD. To probe this transport mechanism, we
plot the probability density and electron current density in real
space shown in \fig{V023X1M}.  Without a central barrier, the
electrons are easily coupled to the LD and then doing cyclotron
motion coupled backward with alignment to the second quasibound
state in the UD causing a resonant transmission [see
\fig{V023X1M}(a)].  For $V_2 = 1.0$ meV such an alignment in energy
disappears, as a result no hump structure can be found. When the
strength of the central barrier increases [see
\fig{V023X1M}(b)-(c)], the electron energy is gradually allowed to
align the first quasibound state energy in the UD  to facilitate a
resonant transmission.  It is interesting to mention that for the
case of $V_2 =0$ meV the shoulder of the hump structure at $X =
1.58$ corresponds to a resonant transmission with lateral tunneling
through the central barrier weakly coupling to the LD as displayed
in \fig{V0X1B5}(a).

Now we turn to discuss the high-KE single-mode regime.  The general
transport characteristics are small peaks in $G$ with height around
$0.2 G_0$. In this regime, electrons are allowed to strongly couple
to the LD.  For the case without a central barrier, the electrons at
energy $X=1.83$ form a (1,2)-like quasibound state in the parallel
DOQD system as is shown in \fig{V0X1B5}(b).  The second small peak
($X=1.97$) in $G$ corresponds to a mixed (2,2)-like quasibound state
[see \fig{V0X1B5}(c)].  This state is constructed by four localized
electronic cyclotron orbits---two stronger in front of and behind
the central barrier and two weaker around the UD and the LD---to
form a {\it hole-like quasibound state} at the center of the DOQD
system.
\begin{figure}[t]
      \includegraphics[width=0.45\textwidth,angle=0]{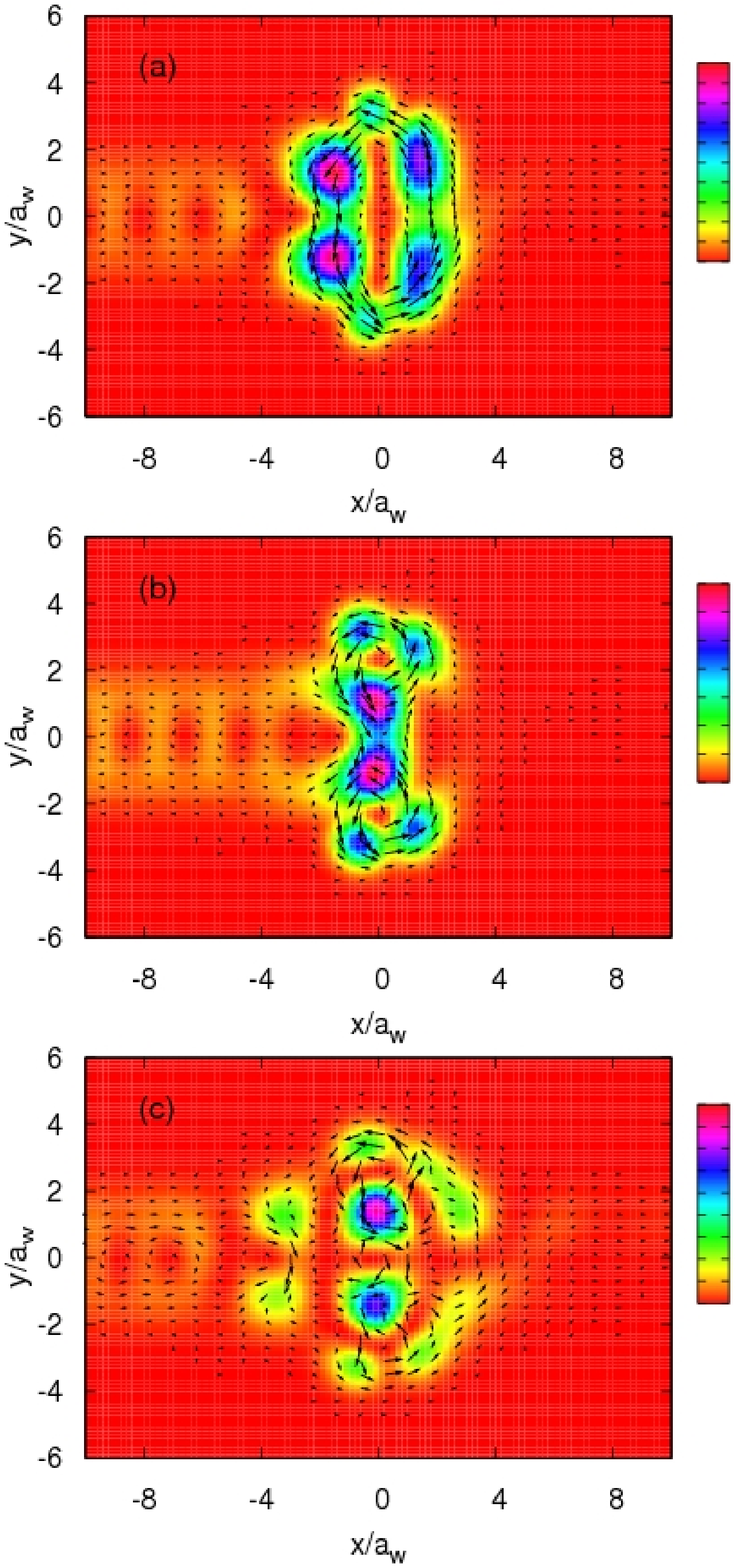}
      \caption{Probability density and electron current density (black arrows)
         for the case of $V_2 = 2.0$ meV and $B = 0.5$ T at $X =$ (a) 1.68; (b) 1.80;
         and (c) 1.91. Other parameters are the same as \fig{V0X1B5}.
               } \label{V2X1B5}
\end{figure}

If we increase the height of the central barrier to $V_2 = 2.0$ meV,
we find that there are three small resonant peaks in $G$ at
$X=1.68$, $1.80$, and $1.91$. Their corresponding transport patterns
in real space are shown in \fig{V2X1B5}.  The first small peak at
$X=1.68$ is a perfect (2,2)-like quasibound state~\cite{Tang03} as
demonstrated in \fig{V2X1B5}(a).  We note that there are two valley
structures, on the sides of this resonant peak, with conductance
minima at $X=1.66$ and $1.76$. At these incident energies, the
electrons only have good coupling to the UD due to the Lorentz force
induced shifting effect.  When the electrons are well-coupled to the
whole DOQD system, such shifting effect becomes unimportant and the
transport feature exhibits a clear resonant peak in $G$.

In \fig{V2X1B5}(b), we show the electronic transport behavior at the
second small peak ($X=1.80$). It is clearly seen that the
counterclockwise cyclotron orbits are separated symmetrically by the
central barrier.  Due to the Lorentz force, the electrons are able
to laterally tunnel through the central barrier spending more time
on each side of the barrier thus forming a (1,2)-like quasibound
state.  With higher incident energy $X=1.91$, not only the tunneling
feature of the UD-LD coupling is present, the electrons are also
able to do cyclotron motion bypassing the central barrier [see
\fig{V2X1B5}(c)]. It is interesting to mention in passing that the
small peaks at $X= 1.61$, $1.71$, and $1.87$ for the case of $V_2 =
1.0$ meV as well as the small peaks at $X=1.73$, $1.86$, and $1.94$
for the case of $V_2 = 3.0$ meV have transport features similar to
those of $V_2 = 2.0$ meV.

\begin{figure}[t]
      \includegraphics[width=0.45\textwidth,angle=0]{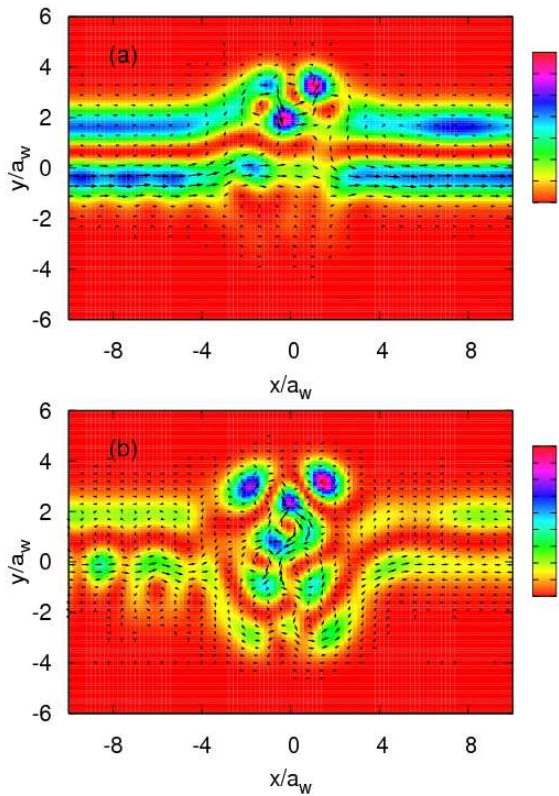}
      \caption{Probability density and electron current density (black arrows)
         for the case of $V_2 = 0$ meV and $B = 0.5$ T at $X =$ (a) 2.28 and (b) 2.51.
         Other parameters are $V_1=V_3=-6.0$ meV, $L=4.82 a_w$, and
         incident mode $n=1$.
               } \label{V0X2MB5}
\end{figure}

In contrast to the single-mode regime, one can find a conductance
peak in the low-KE double-mode regime only for the case of $V_2 = 0$
meV at $X=2.01$ [see the solid curve in \fig{GEV}(a)].  This peak
structure indicates a resonant transmission with coupling to the
lowest quasibound state in the UD.  For this case without a central
barrier and in the mediate-KE double-mode regime, there are two
small peak structures in $G$. First, a resonant transmission due to
a magnetic Fabry-P{\'e}rot-type resonance in the upper open quantum
dot in the upper incident channel is found at $X=2.28$ forming a
resonant peak structure in $G$ as is clearly demonstrated in
\fig{V0X2MB5}(a).  The electrons in the lower channel transport
directly through the DOQD system with coupling to the upper channel.
On the other hand, electrons with incident energy $X=2.51$ manifests
a smaller peak structure in $G$ that implies a short-lived resonant
state. Figure \ref{V0X2MB5}(b) shows that the transport mechanism of
this small peak structure is nontrivial: The electrons can form a
(2,1)-like quasibound state in the UD with {\it magnetic turbulence}
in the central part of the wire and weakly coupling to the LD. A
more complicated magnetic turbulence feature can be found if
electrons are incident from the first mode ($n=0$), due to their
higher electron kinetic energy.

\begin{figure}[t]
      \includegraphics[width=0.45\textwidth,angle=0]{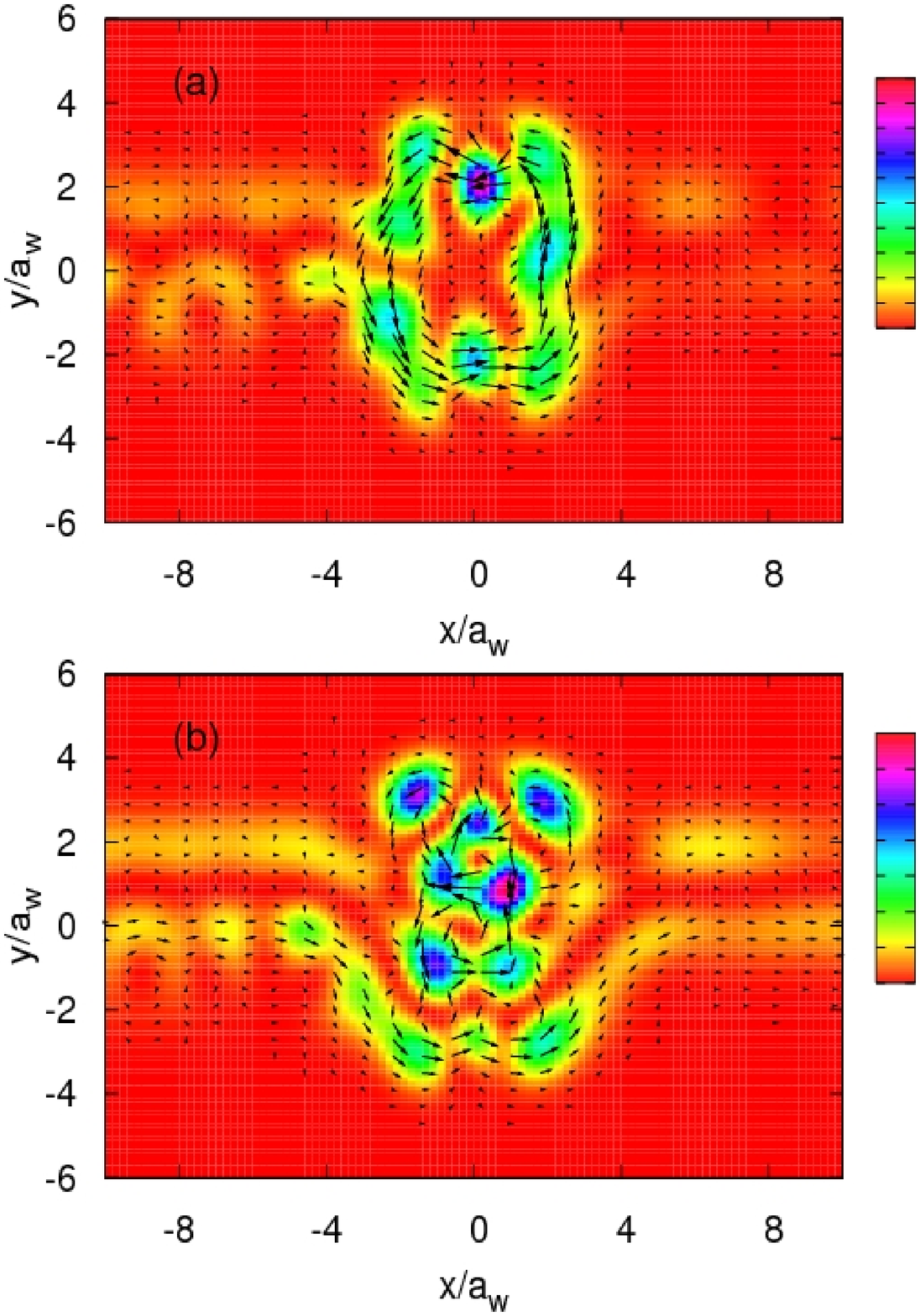}
      \caption{Probability density and electron current density (black arrows)
         for the case of $V_2 = 1.0$ meV and $B = 0.5$ T at $X =$ (a) 2.31 and (b) 2.56.
         Other parameters are the same as \fig{V0X2MB5}.
               } \label{V1X2MB5}
\end{figure}
Contrarily, for $V_2=1.0$ meV in the mediate-KE double-mode regime,
there are two small dip structures in $G$ at $X= 2.31$ and $2.56$.
First, the electrons with incident energy $X=2.31$ prefer to do a
round trip motion around the central barrier forming a strong UD-LD
coupling to construct a perfect quasibound state in the parallel
DOQD system as depicted in \fig{V1X2MB5}(a).  In addition, a weaker
peak is found at $X=2.56$ indicating a shorter living quasibound
state. Indeed, as shown in \fig{V1X2MB5}(b), both the probability
density and the electron current density are complicated indicating
a mixed resonant feature in transport.  A resonant state is formed
at the central part of the DOQD system with stronger coupling to the
UD and weaker coupling to the LD due to the applied magnetic field
in ${\hat{\bf z}}$ direction.

\begin{figure}[t]
      \includegraphics[width=0.45\textwidth,angle=0]{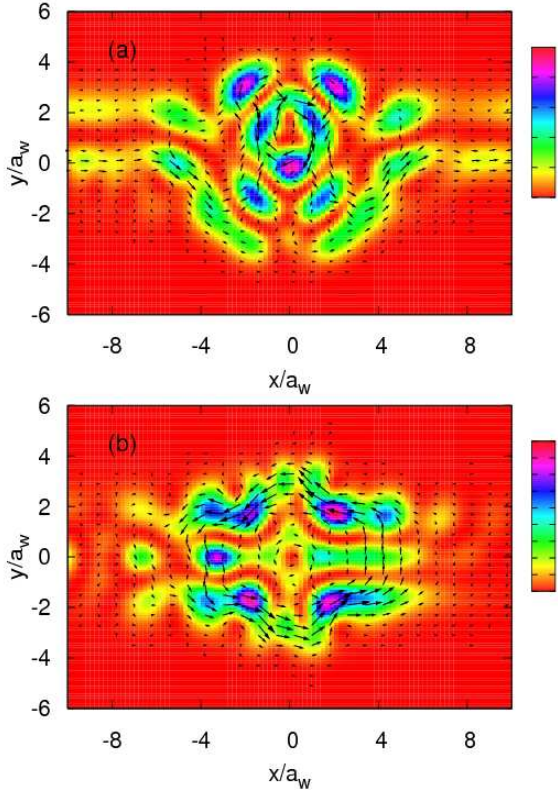}
      \caption{Probability density and electron current density (black arrows)
         for the case of $V_2 = 1.0$ meV in the high-KE double-mode regime
         in a magnetic field $B = 0.5$ T at $X =$ (a) 2.86 and (b) 2.96.
         Other parameters are $V_1=V_3=-6.0$ meV, $L=4.82 a_w$, and
         incident mode $n=1$.
               } \label{V1X2HB5}
\end{figure}
Now we turn to discuss the case of $V_2 =1.0$ meV in the high-KE
double-mode regime.  There are two resonant features in $G$: A small
sharp peak at $X=2.86$ and a deep sharp dip at $X=2.96$ [see the
dashed curve in \fig{GEV}(a)].  The sharpness of the two structures
implies a pure and long-lived resonant state.  First, electrons with
incident energy $X=2.86$ are able to construct a hole-like resonant
state at the upper central part of the parallel DOQD system as
illustrated in \fig{V1X2HB5}(a). Specifically, the electrons have
two channels incident from the left lead, that can either couple to
the LD or being backscattered to the UD. Those electrons transported
through the LD can turn back due to the cyclotron motion and then
couple to the UD. Again, electrons may be backscattered by the UD
and then coupled to the central barrier.  As such, the electrons
construct a {\it magnetic hole-like state} in the DOQD system.
Second, the sharp dip in $G$ at $X=2.96$ implies that a long-lived
pure quasibound state is formed in the DOQD system.  Indeed, as
shown in \fig{V1X2HB5}(b), the probability density shows a
(2,3)-like quasibound state constructed in the system, and the weak
link to the left and the right lead indicates the long dwell time of
this resonant state. Moreover, the electron current density forms a
perfect circular motion around the parallel DOQD indicating a pure
quasibound state.

Correspondingly, for the case of $V_2 =0$ meV in the high-KE
double-mode regime, there are two structures in $G$: a shallow dip
at $X=2.78$ and a deep sharp dip at $X=2.93$ [see the solid curve in
\fig{GEV}].  The corresponding probability density and the electron
current density distribution of these two resonance states are
similar to the case of $V_2 = 1.0$ meV in the high-KE double-mode
regime.  However, unlike the small sharp peak in $G$ at $X=2.86$ for
the case of $V_2 = 1.0$ meV, electron transport with shallow dip in
$G$ at $X=2.78$ for the case of $V_2 = 0$ meV is not able to form a
hole-like bound state due to the absence of backscattering from the
central barrier. Instead, electron can form a quasibound state at
the center of the parallel DOQD system.  The sharp dip in $G$ at
$X=2.93$ corresponds to a pure (2,3)-like quasibound state, which is
similar to the dip structure at $X=2.96$ for the case of $V_2 = 1.0$
meV.  The similarity of the transport features is due to the large
loop of the symmetric cyclotron motion.  Therefore, the central
barrier plays an insignificant role for such a transport feature.
Furthermore, the three probability density peaks in the transverse
direction imply that this (2,3)-like quasibound state is caused by
the vicinity of the third subband threshold in the leads.
\begin{figure}[t]
      \includegraphics[width=0.45\textwidth,angle=0]{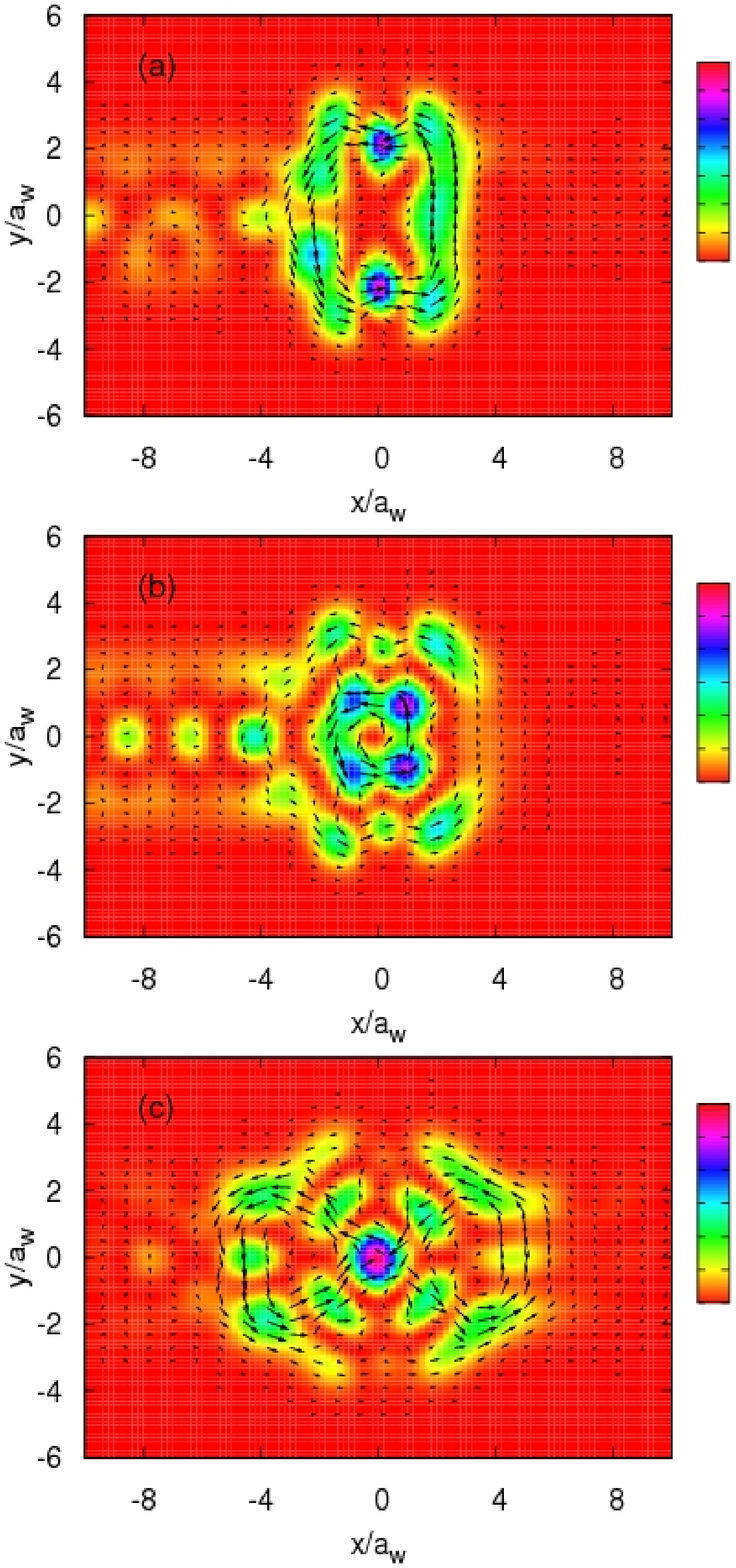}
      \caption{Probability density and electron current density (black arrows)
         for the case of $V_2 = 2.0$ meV in the double-mode regime at $X =$ (a) 2.35;
         (b) 2.63; and (c) 2.94.  Other parameters are $B = 0.5$ T,
         $V_1=V_3=-6.0$ meV, $L=4.82 a_w$, and incident mode $n=1$.
               } \label{V2X2B5}
\end{figure}

We now discuss the case of $V_2=2.0$ meV in the double-mode regime,
there are three significant deep and sharp dips in $G$ as displayed
by the solid curve in \fig{GEV}(b).  The first two sharp dips are at
$X=2.35$ and $2.63$ in the mediate-KE regime.  First, for the case
of the dip at $X=2.35$, the electrons make a perfect (2,1)-like
quasibound state with clear cyclotron motion as demonstrated in
\fig{V2X2B5}(a).  By tuning the central barrier and the energy of
the incident electron to this condition, the UD and the LD can be
strongly coupled with a long life time.  Second, for the case of dip
at $X=2.63$, the electron probability density forms a double ring
structure in real space, and the electron current flows
counterclockwise following this double ring loop with interference
at the UD and the LD regions as is shown in \fig{V2X2B5}(b).  Such a
resonant state formed around the central barrier is a quasibound
state with negative binding energy $-0.91$ meV.~\cite{Vidar05}
Turning to the very narrow and deep dip at $X=2.94$ in the high-KE
double-mode regime, the electrons construct a very clear long-living
state at the origin in real space as depicted in \fig{V2X2B5}(c).
The transport mechanism is that the electrons make intersubband
transitions to the subband top at the origin of the central barrier
forming a quasibound state.  Two clear cyclotron orbits are seen in
front of and behind the central barrier [see \fig{V2X2B5}(c)].

\subsection{Tuning the magnetic field}

Before we discuss the tuning effects on the coherent quantum
transport by adjusting the strength of the external perpendicular
magnetic field, we would like to mention that due to the complex
potential-envelop nature of the laterally parallel DOQD system,
there are several relatively short length scales of the system
leading to the enhanced sensitivity to magnetic field in the range
$0-1$ Tesla.

\begin{figure}[t]
      \includegraphics[width=0.48\textwidth,angle=0]{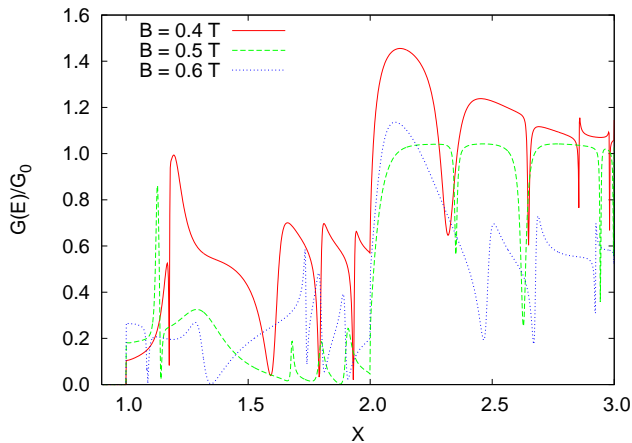}
      \caption{The conductance in units of $G_0=2e^2/h$
               as a function of $X$ for a DOQD system with central barrier
               $V_2=2.0$ meV in a magnetic field $B =$ (a) 0.4 T (solid);
               (b) 0.5 T (dashed); and (c) 0.6 T (dotted).
               Other parameters are $V_1=V_3=-6.0$ meV and $L\simeq 141$ nm.
               } \label{GEB}
\end{figure}

Figure \ref{GEB} shows the magnetic field effects on the conductance
to the DOQD system for an appropriate central barrier $V_2 = 2.0$
meV. The strength of the magnetic field is chosen to be $B=0.4$,
$0.5$, and $0.6$ T, which correspond, respectively, to the effective
magnetic length $a_w=30.59$, $29.34$, and $29.10$ nm; the effective
subband separation $\hbar\Omega_w = 1.22$, $1.32$, and $1.44$ meV;
and the effective length of the DOQD system is $L/a_w=4.62$, $4.82$,
and $4.86$.  It is clearly shown in \fig{GEB} that, in the
single-mode regime, the magnetic blocking effect is most significant
for the case of $B=0.5$ T.  On the other hand, in the double-mode
regime, the general suppressed plateau feature in the conductance is
changed for the case of $B=0.4$ and $0.6$ T: enhanced in the low-KE
regime and suppressed in the high-KE regime.

\begin{figure}[t]
      \includegraphics[width=0.45\textwidth,angle=0]{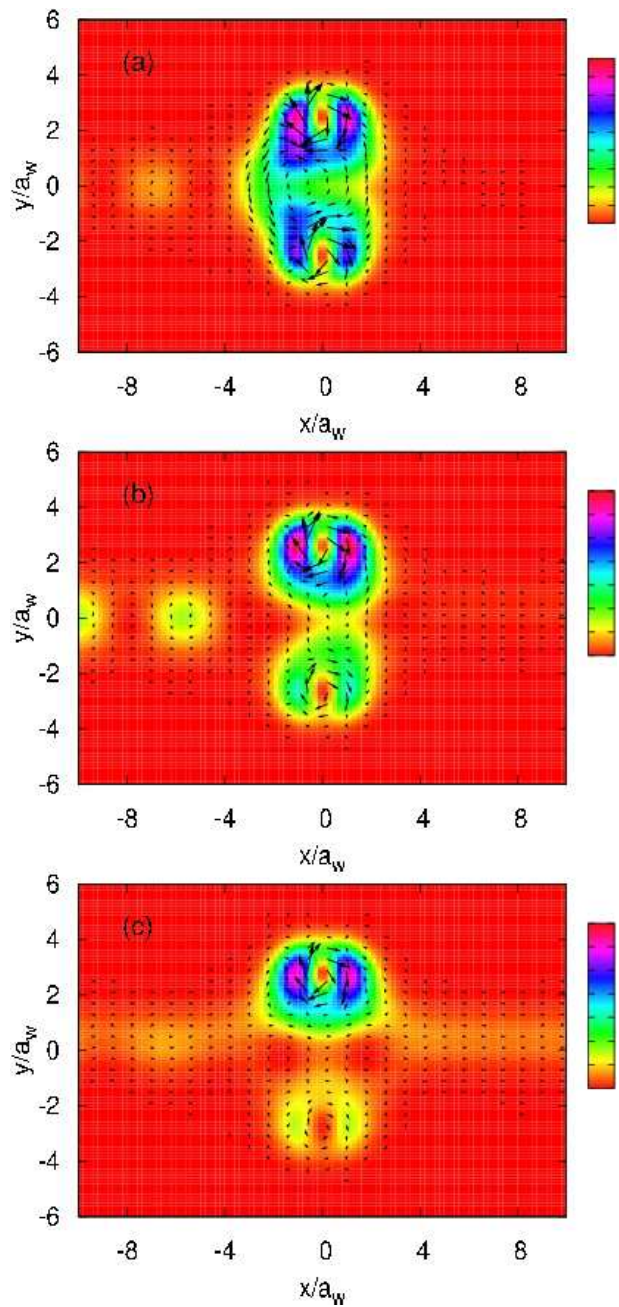}
      \caption{Probability density and electron current density (black
               arrows): (a) at $X=1.18$ for $B=0.4$ T; (b) at $X=1.14$ for $B=0.5$ T;
               and (c) at $X=1.09$ for $B=0.6$ T.
               Other parameters are $V_2 = 2.0$ meV, $V_1=V_3=-6.0$ meV,
               and $L\simeq 141$ nm.
               } \label{B456X1L}
\end{figure}

In the low-KE single-mode regime, there are three sharp downward
dips in $G$ at $X=1.18$ for $B=0.4$ T (solid), $X=1.14$ for $B=0.5$
T (dashed), and $X=1.09$ for $B=0.6$ T (dotted).  The transport
mechanism of these three dips is similar: The incident electron
energy has good alignment to the discrete levels in the upper open
dot forming a magnetic hole-like (2,1) quasibound state.  For a
weaker magnetic field $B=0.4$ T,  the electrons with larger
cyclotron orbit can bypass the central barrier, and hence the UD-LD
coupling is strong enough to form a symmetric probability density
pattern as is shown in \fig{B456X1L}(a).  Therefore, the conductance
manifests a very sharp downward dip in response to this robust
effect.  When the magnetic field is further increased, the hole-like
quasibound state formed in the lower open dot is getting still
weaker. This trend is clearly demonstrated in \fig{B456X1L}(b)-(c).
For the case of $B=0.5$ T, the mediate UD-LD coupling leads to a
clear Fano structure in the conductance [see the dashed curve in
\fig{GEB}]. However, when the magnetic field is increased to $B=0.6$
T, the UD-LD coupling is further weaker leading to vanishing Fano
structure and, instead, the electron conduction manifests a simple
dip structure with zero conductance as depicted by the dotted curve
in \fig{GEB}.

\begin{figure}[t]
      \includegraphics[width=0.45\textwidth,angle=0]{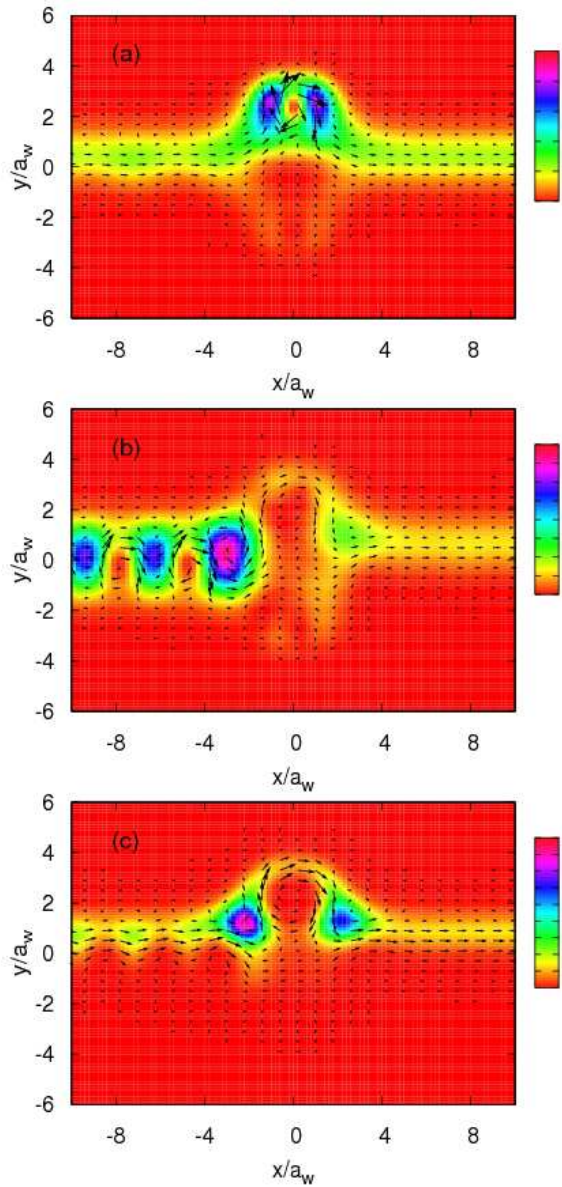}
      \caption{Probability density and electron current density (black
      arrows): (a) at $X=1.20$ for $B=0.4$ T;
               (b) at $X=1.29$ for $B=0.5$ T; and (c) at $X=1.35$ for $B=0.6$ T.
               Other parameters are $V_2 = 2.0$ meV, $V_1=V_3=-6.0$ meV.
               } \label{B456X1M}
\end{figure}

Turning to the mediate-KE single-mode regime, there are three
significant resonant structures in $G$ [see \fig{GEB}]: The peak
structure at $X=1.20$ for $B=0.4$ T, the hump structure at $X=1.29$
for $B=0.5$ T, and the valley structure at $X=1.35$ for $B=0.6$ T.
First, the transport mechanism of the peak structure (for $B=0.4$ T)
is related to the dip structure shown in \fig{B456X1M}(a) forming a
Fano-like line shape.  This peak structure is due to electrons with
strong coupling to the upper open dot forming a hole-like (2,1)
quasibound state and making resonant transmission [see
\fig{B456X1M}(a)]. Unlike the dip structure at $X=1.18$, there is
almost no coupling to the lower open dot thus facilitating electron
conduction, as coupling to the LD usually enhances backscattering to
the left lead. Second, concerning the hump structure for the case of
$B=0.5$ T, the energy alignment to the upper open dot is not
present. Hence, the electrons are not easily coupled to the parallel
DOQD system.  Instead, the electrons form a short-lived quasibound
state in front of the central barrier as illustrated with
\fig{B456X1M}(b). Third, for the case of valley structure at
$X=1.35$ for $B=0.6$ T, an electron incident from the left reservoir
impinges on the central barrier and could either couple to the UD or
being transported through the lower part bypassing the central
barrier forming a resonance state on the left edge of the system, as
is shown in \fig{B456X1M}(c).  The multiple cyclotron scattering
between the two quasibound states, formed on the two edges of the
central barrier, may backward tunnel through the central barrier
exhibiting a zero conductance at $X=1.35$ [see the dotted curve in
\fig{GEB}].

It is interesting to mention in passing that, in the mediate-KE
single mode-regime, a hole-like quasibound state is constructed in
the upper open quantum dot for the case of $B=0.4$ T.  When the
magnetic field is increased to $0.5$ Tesla, such a hole-like state
has vanished. However, for a stronger magnetic field $B=0.6$ T, the
hole-like quasibound state appears again with significant magnetic
multiple scattering on the two edges of the parallel DOQD system as
is clearly demonstrated in \fig{B456X1M}.

\begin{figure}[t]
      \includegraphics[width=0.45\textwidth,angle=0]{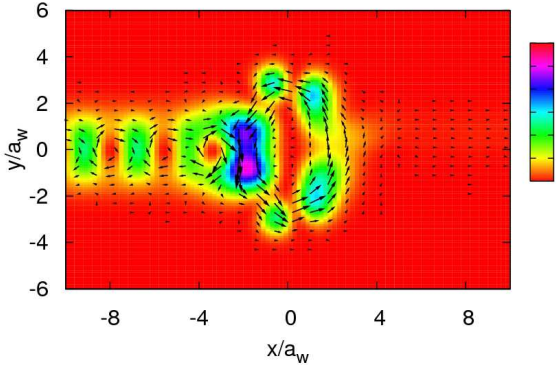}
      \caption{Probability density and electron current density (black
      arrows) for the case of $B=0.4$ T at $X=1.59$.
               Other parameters are $V_2 = 2.0$ meV and $V_1=V_3=-6.0$ meV.
               } \label{V2X1B4}
\end{figure}

For the case of $B=0.4$ T in the higher kinetic energy single-mode
regime, the conductance manifests strong oscillation effects as
shown by the solid curve of \fig{GEB}.  A broad and deep dip
structure is found at $X=1.59$, the electrons incident at this
energy are scattered by the central barrier but can perform a big
cyclotron orbit through the system with UD-LD coupling and construct
a (2,1)-like quasibound state on the right edge of the DOQD system,
as demonstrated in \fig{V2X1B4}. On the other hand, those electrons
that are backscattered by the central barrier may interplay with the
incident electrons constructing a hole-like quasibound state in
front of the central barrier.  In \fig{GEB}, the sharp downward dips
at $X=1.79$ and $1.93$ for the case of $B=0.4$ T have similar
probability and electron current density patterns to, respectively,
the small peaks at $X=1.80$ and $1.91$ for the case of $B=0.5$ T
[see \fig{V2X1B5}(b)-(c)].  When the magnetic blocking effect is
significant, the electronic transport manifests resonant
transmission.  However, if the magnetic blocking effect is
insignificant, the transport feature tends to be resonant
reflection.
\begin{figure}[t]
      \includegraphics[width=0.45\textwidth,angle=0]{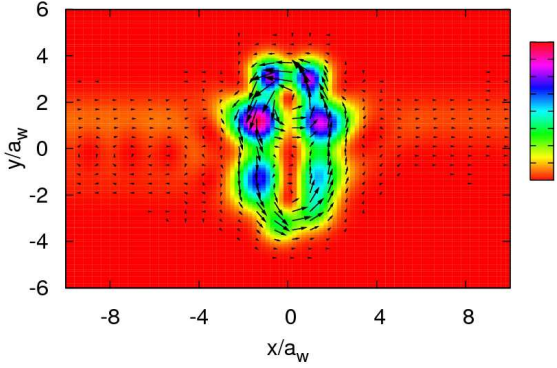}
      \caption{Probability density and electron current density (black
      arrows) for the case of $B=0.6$ T at $X=1.73$.
               Other parameters are $V_2 = 2.0$ meV and $V_1=V_3=-6.0$ meV.
               } \label{V2X1B6}
\end{figure}

For a stronger magnetic field, $B=0.6$ T, the conductance manifests
three Fano line-shapes in the high-KE single-mode regime. The
transport features for the Fano dips at $X=1.74$, $1.81$, and $1.91$
are very similar to, respectively, the three small dips at $X=1.68$,
$1.80$, and $1.91$ for the case of $B=0.5$ T as is shown in
\fig{V2X1B5}.  For the case of $B=0.6$ T, the transport features of
the Fano peaks at $X=1.79$ and $1.89$ are similar to their Fano dips
but with stronger coupling to the right lead. However, the transport
feature of the first Fano peak at $X=1.73$ has significant
difference with its corresponding Fano dip at $X=1.74$. The
probability density and the electron current density pattern of the
Fano peak is shown in \fig{V2X1B6}.  The electrons construct a
(2,2)-like quasibound state between the UD and the central barrier.
In addition, the electrons can also couple to the LD forming a
complete cyclotron motion in the whole DOQD system. The weak
coupling to the two leads shown in \fig{V2X1B6} implies the long
dwell time of this quasibound state.
\begin{figure}[t]
      \includegraphics[width=0.45\textwidth,angle=0]{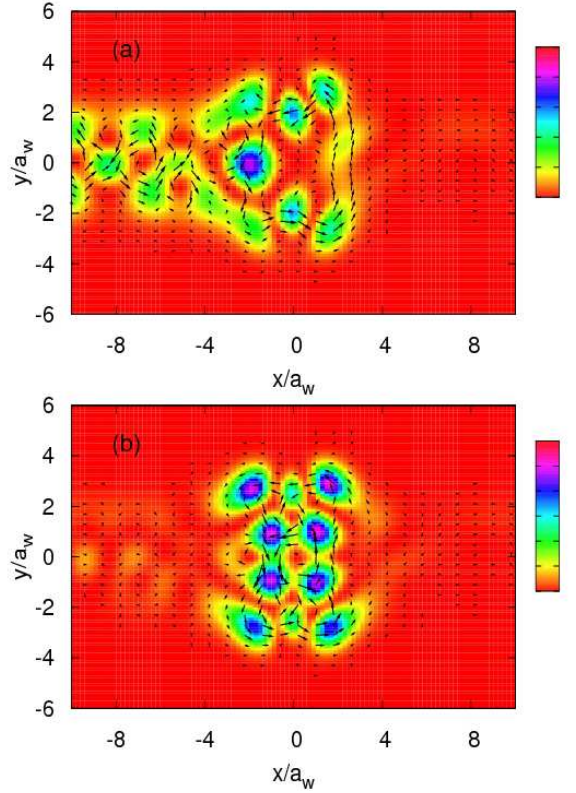}
      \caption{Probability density and electron current density (black
       arrows) for the case of $B=0.4$ T in the mediate-KE double-mode regime
       at $X=$ (a) $2.32$ and (b) $2.65$.  Other parameters are $V_2 = 2.0$ meV,
       $V_1=V_3=-6.0$ meV, and incident mode $n=1$.
               } \label{V2X2MB4}
\end{figure}

Now we turn to study the transport features in the double-mode
regime. In this regime, the conductance manifests significant
downward-dip structures implying resonant reflection. For the case
of $B=0.4$ T in the mediate-KE regime, there are two significant dip
structures in $G$ at $X=2.32$ and $2.55$ as shown by the solid curve
in \fig{GEB}. When the electrons are incident with energy $X=2.32$,
the wave function scars are found not only in the DOQD but also in
the left lead as is shown in \fig{V2X2MB4}(a), this indicates that a
strong quantum interference occurs between two propagating channels
in the left lead.  Moreover, this quantum interference occurring in
the lead also implies a metastable quasibound state with a short
dwell time, and hence the conductance manifests a broad-dip feature.
Furthermore, a strong dot-lead coupling can be found by the high
probability density on the left edge of the DOQD system. On the
other hand, when the electrons are incident with energy $X=2.65$, as
is clearly demonstrated in \fig{V2X2MB4}(b), they can construct a
perfect (2,4)-like quasibound state in the parallel DOQD system.  In
addition, this (2,4)-like state can easily couple to the upper and
the lower open dot thus accomplishing a highly symmetric probability
pattern shown in \fig{V2X2MB4}(b).

\begin{figure}[t]
      \includegraphics[width=0.45\textwidth,angle=0]{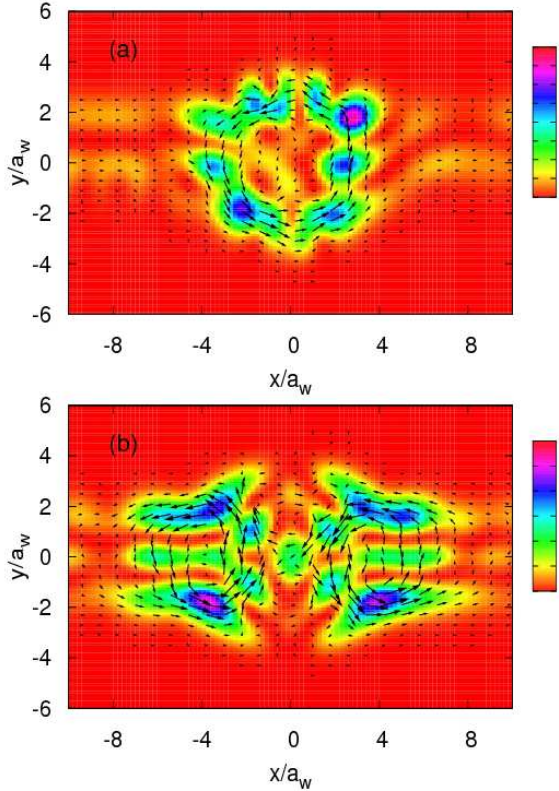}
      \caption{Probability density and electron current density (black
       arrows) for the case of $B=0.4$ T in the high-KE double-mode regime
       at $X=$ (a) $2.85$ and (b) $2.98$.
       Other parameters are $V_2 = 2.0$ meV, $V_1=V_3=-6.0$ meV,
       and incident mode $n=1$.
               } \label{V2X2HB4}
\end{figure}

For the case of $B=0.4$ T in the high-KE double-mode regime, there
are two narrow sharp dips in $G$ at $X=2.85$ and $2.98$ [see the
solid curve in \fig{GEB}]. When the electrons are incident with
energy $X=2.85$,  the electrons have a strong coupling to the
parallel DOQD system forming a (2,3)-like quasibound state, and this
state has weak coupling to the UD and the LD, as depicted in
\fig{V2X2HB4}(a). If the electrons are incident with energy
$X=2.98$, the conductance feature is a very sharp and narrow dip.
The electrons can perform intersubband transitions to the threshold
of the third subband forming a quasibound state in the left and the
right lead as is clearly demonstrated in \fig{V2X2HB4}(b).  The two
clear cyclotron orbits in front of and behind the DOQD system are
found with inversion symmetry implying a long-lived quasibound
state.  Due to the Lorentz force, such a state formed in the lead
may flow back to the system forming (1,2)-like metastable quasibound
states either in front of or behind the central barrier, and these
two metastable states may have weak coupling through the central
barrier.

\begin{figure}[t]
      \includegraphics[width=0.45\textwidth,angle=0]{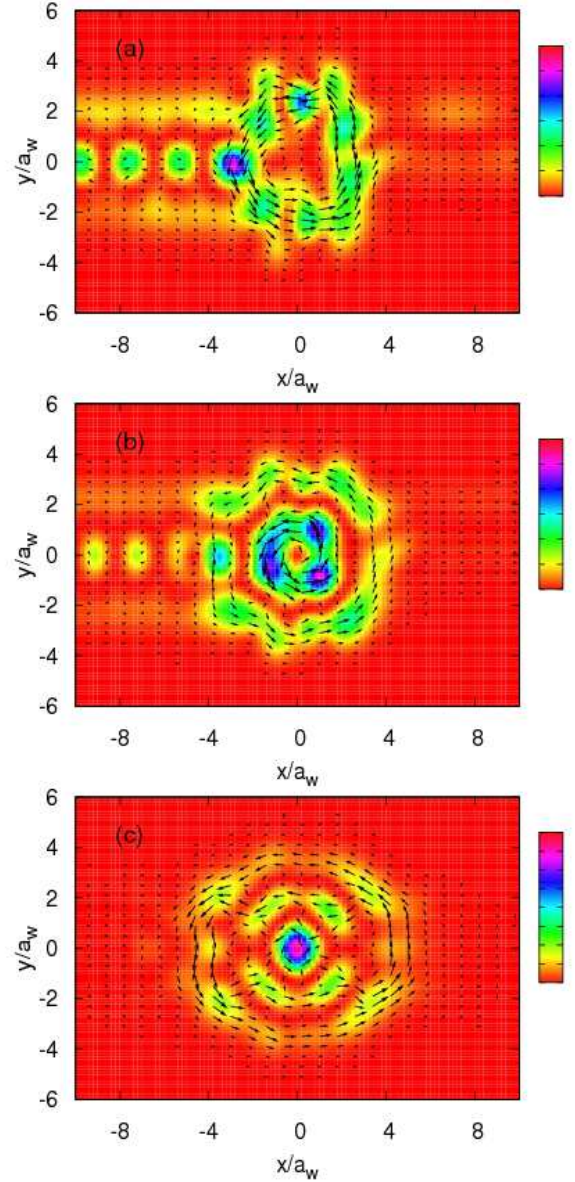}
      \caption{Probability density and electron current density (black
       arrows) for the case of $B=0.6$ T in the double-mode regime
       at $X=$ (a) $2.47$; (b) $2.67$; and (c) $2.92$.
       Other parameters are $V_2 = 2.0$ meV, $V_1=V_3=-6.0$ meV,
       and incident mode $n=1$.
               } \label{V2X2B6}
\end{figure}

For a higher magnetic field ($B=0.6$ T) in the double-mode regime,
the transport characteristics exhibit asymmetric dip features in the
conductance as is shown by the dotted curve in \fig{GEB}.  These
dips are at $X=2.47$, $2.67$, and $2.92$ with an opening line-shape
in the low-energy part of the dip structure and with an abrupt
change in $G$ in the high-energy part of the dip structure.  The
transport features of the three significant asymmetric dips are
displayed in \fig{V2X2B6}(a)-(c).  The general transport features of
these asymmetric dips are related to the ``quantum peg"
effect.~\cite{Vidar05}

The electrons with incident energy $X=2.47$, perform a
cyclotron-type multiple scattering in the DOQD system with strong
constructive quantum interference with incident electron waves on
the left edge of the system as is shown in \fig{V2X2B6}(a). For a
higher incident electron energy $X=2.67$, the cyclotron-type
quasibound state has a longer dwell time and the probability pattern
displays a double-ring structure. It is clearly demonstrated in
\fig{V2X2B6}(b) that a ``quantum peg" structure in real-space
probability density is constructed around the central barrier (a
quasibound state with negative binding energy).  When the incident
electron energy is further raised to $X=2.92$, one can find not only
the double-ring structure but also a probability density peak at the
origin of the central barrier as is depicted in \fig{V2X2B6}(c).

The electron current flows dominantly through the outer ring-shaped
path indicating a perfect quasibound state constructed around the
whole system with long life time.  When the electrons flow through
the outer ring-shape path and close to the UD or the LD, they may
couple to the dot or turn to couple to the inner ring-shaped state.
The low probability density in the UD and the LD implies the
electrons traverse directly through the dots.  On the other hand,
those electrons coupled to the inner path may make intersubband
transitions to a localized subband top formed at the origin of the
central barrier trapped temporarily forming a quasibound state. This
feature is an indirect intersubband transition mechanism similar to
the case of $B=0.5$ T at $X=2.94$ as is shown in \fig{V2X2B5}(c).

It is also worth to point out that for the case of $B=0.6$ T in the
double-mode regime, the asymmetric dip structure at $X=2.92$ is
narrower then the other two dips at lower energies, which in reality
forms a Fano-type line shape.  The Fano-peak at $X=2.93$ has
probability density and electron current density pattern similar to
its corresponding Fano-dip at $X=2.92$, but with stronger coupling
to the right lead. The Fano effect is found to be prominent at the
same specific magnetic fields such as $B=0.6$ T, while it is not
distinct in the other values.  This implies that the coherence
nature of the quantum transport through the laterally parallel DOQD
system strongly depends on the magnetic field.

%
%
\section{conclusions}

We have demonstrated tunable transport effects of coupling modes in
laterally parallel double open quantum dots.  In general the applied
perpendicular homogeneous magnetic field plays a blocking effect on
the quantum transport through the parallel DOQD system. Due to the
complex potential-envelop nature of the system, there are several
relatively short length scales leading to the sensitivity to
magnetic field.

By tuning the central barrier, we have found Fano-type line-shapes
in the conductance in the low-KE single-mode regime.  The peaks of
these Fano-type line-shapes are magnetic hole-like quasibound-state
features in the upper open quantum dot.  In the mediate-KE
single-mode regime, resonant coupling to the upper open dot is
determined by the energy alignment of the incident electron energy
with the quasibound levels in the upper quantum dot.  In the high-KE
single-mode regime, an interesting hole-like quasibound-state
feature has been found in the absence of a central barrier by tuning
the incident electron energy just below the second subband
threshold. Moreover, we have demonstrated the robust features of
tunneling through and bypassing the central barrier in the
transverse direction to achieve the UD-LD coupling.

In the double-mode regime, we have found a magnetic
Fabry-P{\'e}rot-type resonant transmission in the upper open dot if
the incident kinetic energy is low.  In the mediate-KE regime, we
have demonstrated the possibility of the magnetic turbulence effect
in the central part of the system in the absence of a central
barrier.  A magnetic hole-like quasibound state feature has been
found in the high-KE double-mode regime for a weak central barrier.
A clear ``quantum peg" structure can be found in the mediate-KE
double-mode regime for the case of mediate central barrier.
Moreover, in the high-KE double-mode regime, we have found a
quasibound state induced by intersubband transitions to the subband
top located at the origin of the central barrier.

By tuning the magnetic field, we have found a Fano to downward-dip
line-shape crossover on the quantum transport in the low-KE
single-mode regime.  The magnetic field manipulated energy alignment
effect to the upper open dot has also been demonstrated in the
mediate-KE single-mode regime.  In the high-KE single-mode regime,
by increasing the magnetic field from 0.4 Tesla to 0.5 Tesla the
conductance features change from downward dips, small peaks, to Fano
line-shapes.

In the double-mode regime, the effects of wave function scars and a
clear (2,4)-like quasibound state feature have been found for the
case of lower magnetic field in the mediate-KE regime.  For this
lower magnetic field case but with high-KE, the electron transport
manifests a (2,3)-like state and a quasibound state trapped at the
subband threshold in the leads.  When the magnetic field is tuned to
be higher, say 0.6 Tesla, the dip structures in the conductance
become asymmetric and the transport pattern changes to be ``quantum
peg" features.  When the incident energy is just below the third
subband threshold, we can find a robust quasibound state feature
formed at the origin of the central barrier.

In summary, we have proposed a laterally parallel DOQD configuration
for the investigation of tuning effects upon the coupling modes of
lateral parallel quantum dots.  Even for such a simple
configuration, tuning the coupling modes by the central barrier or
by the magnetic field has revealed an essentially complicated nature
of coherent quantum transport.


%
%
\begin{acknowledgments}
      The authors acknowledge the financial support by the Research
      and Instruments Funds of the Icelandic State,
      the Research Fund of the University of Iceland, and the
      Taiwan National Science Council.
      C.S.T. is grateful to the computational facility supported by the
      National Center for High-performance Computing in Taiwan.
\end{acknowledgments}

%
%
\bibliographystyle{apsrev}

%
%
%
\end{document}